%% file: MultigravitonsFinal.tex
\documentclass[a4paper,11pt]{article}
\usepackage[utf8x]{inputenc}

\usepackage{amsmath,amsfonts,amssymb,amsthm,mathrsfs}
\usepackage[vcentermath]{youngtab}
\usepackage{graphicx,graphics}
\graphicspath{{image/}}
\usepackage{cases}
\usepackage{epsfig}
\usepackage{latexsym} % symbols
\usepackage{cite}
\usepackage{hyperref}
\usepackage{xcolor}
\hypersetup{
    colorlinks,
    linkcolor={red!30!black},
    citecolor={blue!40!black},
    urlcolor={blue!80!black}
}
\usepackage{amsmath,amssymb,mathtools}
\usepackage{pdfpages} 
\usepackage{tikz}
 \usepackage{tikz}
\usetikzlibrary{trees,arrows}
\tikzset{level 1/.style={level distance=1.5cm, sibling distance=3.5cm}}
\tikzset{level 2/.style={level distance=1.5cm, sibling distance=2cm}}

\usepackage{pdfpages}
\usepackage[left=2.5cm,top=2.3cm,right=2.5cm]{geometry}

\newsavebox{\uuunit}
\sbox{\uuunit}
    {\setlength{\unitlength}{0.825em}
     \begin{picture}(0.6,0.7)
        \thinlines
        \put(0,0){\line(1,0){0.5}}
        \put(0.15,0){\line(0,1){0.7}}
        \put(0.35,0){\line(0,1){0.8}}
       \multiput(0.3,0.8)(-0.04,-0.02){10}{\rule{0.5pt}{0.5pt}}
     \end {picture}}

\newcommand{\ba}{\begin{eqnarray*}}
\newcommand{\ea}{\end{eqnarray*}}
\newcommand{\ban}{\begin{eqnarray}}
\newcommand{\ean}{\end{eqnarray}}

\def\beq{\begin{equation}}
\def\bee{\begin{equation}}
\def\eeq{\end{equation}}
\def\bea{\begin{eqnarray}}
\def\eea{\end{eqnarray}}
\def\bd{\begin{displaymath}}
\def\ed{\end{displaymath}}
\title{Massive gravitons on the Landscape and the AdS Distance Conjecture}

\author{Ioannis Lavdas$^{\alpha}$\footnote{ioannis.lavdas@physik.uni-muenchen.de}, \quad Dieter L{\"u}st$^{\alpha,\beta}$\footnote{dieter.luest@lmu.de \&  luest@mppmu.mpg.de}}
\date{%
\center{$^{\alpha}$\textit{\small Arnold-Sommerfeld-Center for Theoretical Physics, Ludwig-Maximilians-Universit{\"a}t, 80333 M{\"u}nchen, Germany}}\\
\center{$^{\beta}$\textit{\small Max-Planck-Institut f{\"u}r Physik (Werner-Heisenberg-Institut), F{\"o}hringer Ring 6, 80805, M{\"u}nchen, Germany}}\\[2ex]%
}

\begin{document}

\maketitle

\begin{center}
\textbf{Abstract}
\end{center}
\hfill\break
This work regards IIB string theory embeddings of massive and bi-metric $AdS_{4}$ gravity where the spectrum of the theory includes multiple massive gravitons. The corresponding geometry consists of multiple $AdS_{4}\times_{w}\mathcal{M}_{6}$ spacetimes coupled by Janus throats of different radii.  In the second part of this work, the AdS distance conjecture is studied in the above context and it is shown that the scale of the predicted infinite tower of light states is related to the breakdown scale of the effective field theory.

\vskip-13cm\hskip10cm
{LMU--ASC 32/20, MPP-2020-119}

\newpage 
\tableofcontents
\newpage

\section{Introduction}
 The Swampland program \cite{Vafa:2005ui,Palti:2019pca} has recently achieved a remarkable progress which is grounded on the formulation and the extended applications of the generalised distance conjecture \cite{Ooguri:2006in} and its generalisation to anti-de Sitter space  \cite{Lust:2019zwm}. A close relation of the generalized distance conjecture to the geometric Ricci flow was elucidated in \cite{Kehagias:2019akr}.
The generalized distance conjecture can engage problems regarding the UV completion of effective field theories, which arise in the context of string theory on $AdS_{p}\times \mathcal{M}$ product manifolds.
 Among those, questions regarding the validity range of effective field theories of massive and bimetric gravity  \cite{deRham:2014zqa,Hinterbichler:2011tt,Schmidt-May:2015vnx}
   in anti-de Sitter space are of particular interest, as they have received a lot of attention due to progress that has been achieved from different approaches 
   \cite{deRham:2010kj,ArkaniHamed:2002sp,Chamseddine:2018gqh,Alberte:2019lnd}.
  \hfill\break
 In this work we consider the top-down embeddings of massive and bimetric gravity  on $AdS_{4}$ in type IIB string theory, developed in \cite{Bachas:2017rch,Bachas:2018zmb}. The two main results of these works 
 are the demonstration that massive $AdS_{4}$ gravity belongs to the string theory landscape and the derivation of a
 quasi-universal formula for the mass of the lowest lying spin-2 mode. The effective field theory breakdown scale for massive gravity on $AdS$ was identified in \cite{deRham:2016plk}. A question of high significance is the origin of this breakdown in the context of the aforementioned embeddings. This question is treated in \cite{Bachas:2019rfq} using holography where the origin of the breakdown is the appearance of an infinite tower of spin$\geq2$ modes in the limit $m_{g}\rightarrow 0$. \footnote{A similar discussion about the swampland for massive
 gravity in flat space-time was put forward in \cite{Klaewer:2018yxi}.} In this work a massive-AdS-graviton distance conjecture was formulated, as this limit is reached at infinite distance in the scalar moduli space.  
 Extending this analysis in the case of multiple interacting massive gravitons is the task that is described is the first part of the present work and along with certain special examples and a statement for the breakdown scale in this context, completes the picture of these embeddings. 
 \hfill\break
Furthermore, it should in principle be possible to arrive to derive the breakdown scale of the effective field theory by following the approach of the generalised distance conjecture. The conjecture, when applied to the case of supersymmetric AdS vacua, boils to the AdS Distance conjecture \cite{Lust:2019zwm}, which refers to the appearance of an infinite tower of light modes  as $\Lambda\rightarrow 0$, introducing in this way naturally the scale at which the effective theory should break down. In the second part of this work this question is treated by applying the above statement in our $AdS_{4}$ massive gravity context and by comparing with the universal result of \cite{deRham:2016plk}. This in a sense constitutes on the one hand a threefold approach to the question of the effective theory breakdown scale and on the other hand the formulation of the statement of the generalised distance conjecture by relating the mass scale of the infinite tower with the graviton mass. 

\hfill\break
The paper is organised as follows: In section 2, after a brief review of the aforementioned embeddings , we introduce their extended version, which includes multiple massive gravitons. The corresponding geometries are six manifolds connected by thin (Janus) throats of diverse radii. The wavefunctions of the massive gravitons and their leading order masses are computed for supergravity solutions dual to linear or circular quivers. In section 3, we introduce the swampland distance conjecture and its generalised version, along with its strong version in the case of supersymmetric AdS vacua. Then these conjectures are appropriately formulated along the lines of the problem we are studying and finally the scale of the conjectured tower is compared with the breakdown scale of the effective field theory for massive gravity on $AdS_{4}$.

%%%%%%%%%%%%%%%%%%%%%%%%%%%%%%%%%%%%%%%%%%%%%%%%%%%%%%%%%%%%

%%%%%%%%%%%%%%%%%%%%%%%%%%%%%%%%%%%%%%%%%%%%%%%%%%%%%%%%%%%%
\newpage
\section{Multiple massive $AdS_{4}$ graviton solutions}

The construction of solutions of type IIB string theory with multiple massive (lowest lying) spin-2 modes, follows from the extension of the ones of the recent configurations \cite{Bachas:2017rch,Bachas:2018zmb}. These constitute the first top-down embedding of massive AdS supergravity to string theory, indicating that massive gravity in $AdS_{4}$ belongs to the string Landscape. Prior to the presentation of the multi-graviton solutions, these constructions are briefly presented below for consistency.

\subsection{String theory embeddings of Massive and Bimetric $AdS_{4}$ gravity}

The classical solutions which allow for a small graviton mass are based on the ones initially introduced in \cite{Karch:2000ct,Karch:2001cw,DHoker:2007zhm}. Their general form is an $AdS_{4}$ fibration over a six-dimensional base manifold which is the warped product of a pair of two-spheres over a two-dimensional Riemann surface:

\hfill\break
\begin{equation}
ds^{2}=L_{4}^{2}ds_{\text{AdS}_{4}}^{2}+ds_{\mathcal{M}_{6}}^{2};\quad \mathcal{M}_{6}=S^{2}\times\hat{S}^{2}\ltimes\Sigma_{(2)}
\end{equation}
\hfill\break

The solutions include point singularities on the two boundaries of the strip which are interpreted as five-branes (D5 and NS5) in string theory. The solution is then characterised by the brane charges (NS-NS and R-R three-form flux, R-R five-form flux), as well as from the value of the dilaton field. The solutions are presented in detail in the Appendix.
\hfill\break
The above solutions are dual to a class of three-dimensional superconformal theories with $\mathcal{N}=4$ supersymmetry. They arise as non-trivial infrared fixed points of the renormalisation group flow of theories the field content (gauge groups and matter fields) of which is arranged in terms of linear quiver diagrams \cite{Gaiotto:2008ak,Hanany:1996ie}.  The precise holographic dictionary between the aforementioned supergravity solutions and these theories was developed in \cite{Assel:2011xz}. 
\hfill\break

The question of the presence of a bulk massive graviton is related to the energy-momentum conservation in the dual CFT. The $AdS_{4}$ graviton is dual to the stress tensor of the $\text{sCFT}_{3}$, the scaling dimension of which is holographically related to the graviton mass \cite{Aharony:1999ti}:
\hfill\break
\begin{equation}
m^{2}L_{4}^{2}=\Delta(\Delta-3)\, ,
\end{equation}
\hfill \break
with $L_{4}$ being the $AdS_{4}$ radius and $\Delta$ the scaling dimension of the stress-tensor. The representation of the three-dimensional conformal algebra $\mathfrak{so}(2,3)$ which contains $T_{\alpha\beta}$ and its conformal descendants, is actually  short, due to the conservation of the stress-tensor which results to three null descendant states. The scaling dimension of the conserved stress-tensor does not receive quantum corrections and hence it is canonical: $\Delta=3$. From the above expression, it becomes obvious that a conserved stress tensor of the boundary three-dimensional CFT corresponds to a massless $AdS_{4}$ graviton. Therefore, a slightly massive graviton would correspond to a non-conserved stress-tensor. 

In this case, its scaling dimension would acquire small anomalous dimension \cite{Aharony:2003qf}, $\varepsilon\ll 1$, which would then correspond to a small graviton mass:

\hfill\break
\begin{equation}
m^{2}L_{4}^{2}\sim \mathcal{O}(\epsilon)
\end{equation}
\hfill\break

\textbf{$AdS_{4}$ Bigravity}:
\hfill\break
\hfill\break
 One mechanism through which this scenario can be realised is the one where two initially decoupled three-dimensional theories are coupled weakly while conformal symmetry is preserved\cite{Bachas:2017rch}. This is realised by gauging a common global symmetry. Finally the two intially disjoint quivers are connected by a node which corresponds to a unitary gauge group of low rank. We will be focusing mostly on the gravitational side where the dual picture is a connection of two initially decoupled $AdS_{4}$ universes via a thin throat. 
 \hfill\break
  \begin{center}
\begin{figure}[ht]
\centering
\includegraphics[width=.8\textwidth,scale=0.6,clip=true]{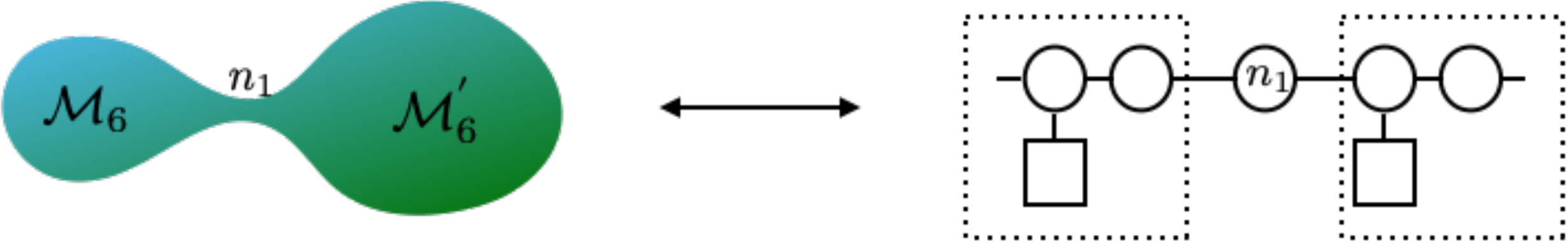}
\caption{Two compact six-manifolds coupled by a highly curved Janus throat and the holographic dual picture, two linear quivers coupled by a low-rank unitary node ($U(n)$).}
\end{figure}
\end{center}
\hfill\break
Initially the spectrum of the theory in each of the two disconnected $AdS{4}\times_{w}\mathcal{M}_{6}$ spacetimes, includes a single massless lowest lying spin-2 mode, since each $\mathcal{M}_{6}$ is initially compact. The coupling of the two geometries is realised in terms of the distance between singularities on the strip that label each theory \cite{Bachas:2017rch} . The geometry of the connecting throat is either $AdS_{5}\times S^{5}$ in the case where the value of the dilaton in the two spacetimes is the same ($\delta\varphi=0$), or its supersymmetric Janus deformation, in the case when the dilaton in the throat interpolates between two constant values in the two spacetimes, determined practically by the number of the five-branes of the solution.
\hfill\break
The eigenvalue problem for the spin-2 spectrum of the theory was considered in detail in \cite{Bachas:2011xa} and is given by the action of the Laplace (mass-squared) operator on the mass eigenstates in the internal manifold, while the norm is provided by the expression for the inner product:
\hfill\break
\begin{equation}
M^{2}\psi=m^{2}L_{4}^{2}\psi; \quad M^{2}:=-\frac{L_{4}^{-2}}{\sqrt{g}}\partial_{i}(L_{4}^{4}\sqrt{g}g^{ij}\partial_{j});\quad \langle\psi_{i}\vert\psi_{j}\rangle:=\int_{\mathcal{M}_{6}}d^{6}y L_{4}^{2}\sqrt{g}\psi_{i}^{*}\psi_{j}
\end{equation}
\hfill\break
By using the explicit expressions for the metric functions of the supergravity solutions we consider, it is straightforward to obtain for the mass eigenvalue:
\hfill\break
\begin{equation}
\langle\psi\vert M^{2}\vert\psi\rangle=\int_{\mathcal{M}_{6}}d^{6}y\sqrt{g}L_{4}^{4}g^{ij}\partial_{i}\psi^{*}\partial_{j}\psi
\end{equation}
\hfill\break
 Once the two spacetimes couple, the two constant wavefunctions mix into two orthogonal combinations: a wavefunction which remains constant all over the new compact manifold and one which varies and changes sign in the throat, interpolating between two constant values in the two $AdS_{4}\times_{w}\mathcal{M}_{6}$ spacetimes. In order to calculate the mass of the lowest-lying graviton, we do not have to solve the full above spectral problem but rather solve for the lowest eigenvalue by minimising the above expression,  with the appropriate boundary conditions. The result for the graviton mass reads:
 \hfill\break
 \begin{equation}
 m_{g}^{2}L_{4}^{2}\sim n^{2}\Big(\frac{\kappa_{4,(1)}^{2}}{\langle L_{4}^{2}\rangle_{\mathcal{M}_{6}^{(1)}}}+\frac{\kappa_{4,(2)}^{2}}{\langle L_{4}^{2}\rangle_{\mathcal{M}_{6}^{(2)}}})\mathcal{J}(\text{cosh}\delta\varphi),
 \end{equation}
 \hfill\break
where $\kappa_{4,(i)}$ is the four-dimensional effective gravitational coupling at each spacetime, $\langle L_{4}^{2}\rangle$ the average squared $AdS_{4}$ radius\footnote{$\langle L_{4}^{2}\rangle=\int\sqrt{g}L_{4}^{2}/\mathcal{V}_{6};\quad \mathcal{V}_{6}=\int\sqrt{g}$}, while $\mathcal{J}(\text{cosh}\delta\varphi)$ is a function of the dilaton variation which trivially contributes to the mass for constant dilaton, while vanishes as the variation of the dilaton grows, suppressing in this way the value of the mass \footnote{An alternative form of the result where the dilaton variation is manifest, was given in \cite{Bachas:2019rfq}: $\varepsilon\sim n^{2}\frac{\text{tanh}^{3}\delta\varphi}{(\delta\varphi-\text{tanh}\delta\varphi)}\frac{1}{c}$, with c being the central charge of the three-dimensional dual CFT.}. Finally, $n$ stands for the quantised $D3$ charge of the throat ($L^{2}_{\text{thr}}\sim n$). The following figure shows the variation of the radius of the $AdS_{4}$ fiber along the throat: it reaches a minimal value inside the throat, where the wave function vanishes exponentially, while approaching constant values in the two coupled universes.
\begin{center}
\begin{figure}[ht]
\centering
\includegraphics[width=.6\textwidth,scale=0.2,clip=true]{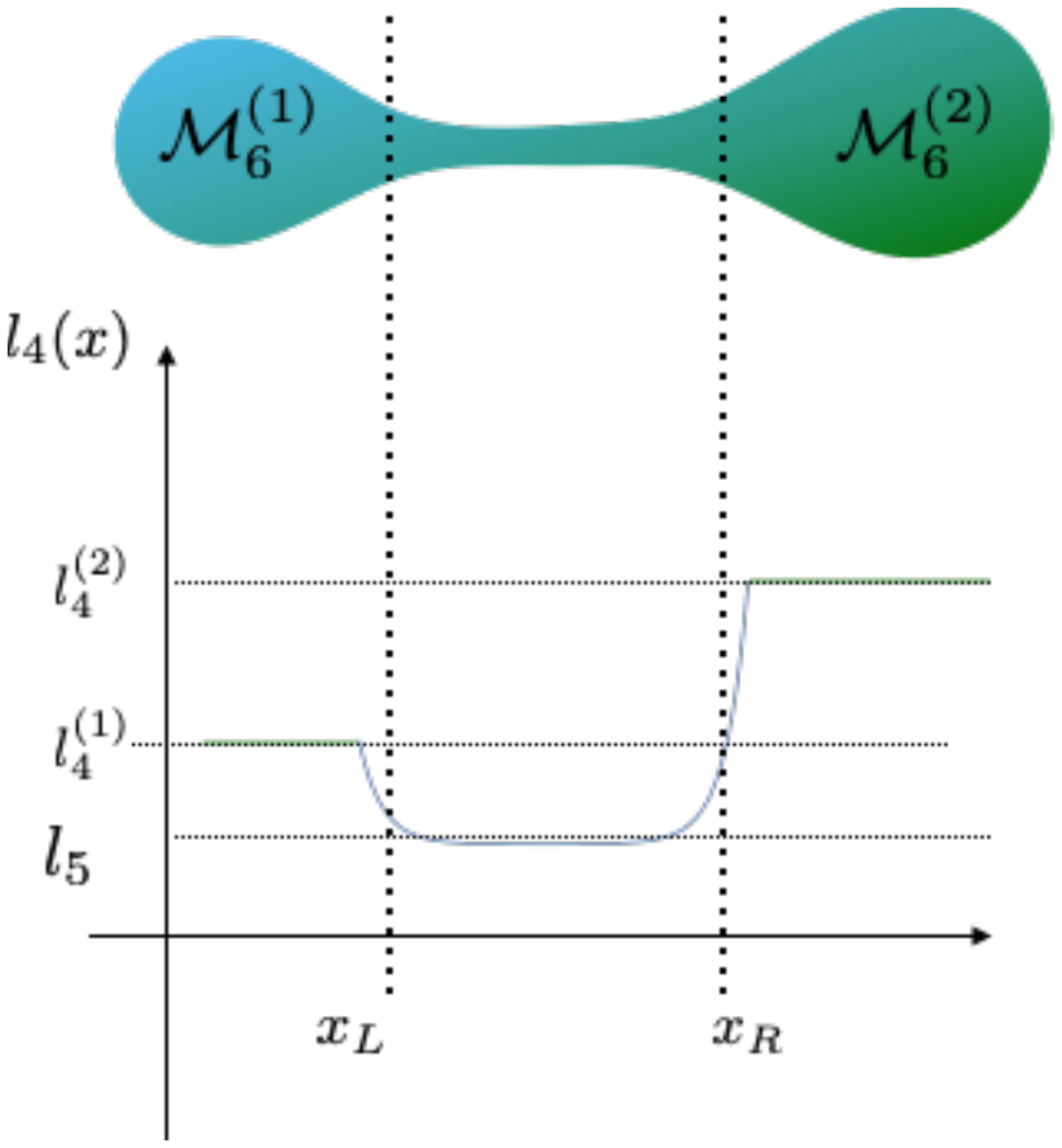}
\caption{Variation of the radius of the $AdS_{4}$ fiber along the geometry.}
\end{figure}
\end{center}
\hfill\break
\hfill\break

\textbf{$AdS_{4}$ massive gravity}
\hfill\break
\hfill\break
A special limit of the above solution is the one where the radius of one of the $AdS_{4}$ spacetimes blows up $L_{4(i)}\rightarrow\infty$ or equivalently when the effective gravitational coupling vanishes ($\kappa_{4(i)}\rightarrow 0$). The spacetime is no longer compact and the geometry of the throat asymptotes $AdS_{5}/\mathbb{Z}_{2}\times S^{5}$ at infinity. As a result, the massless mode decouples as its wavefunciton now vanishes and the spectrum includes only a single massive graviton, the wavefunction of which vanishes at infinity.
\hfill\break
The dual configuration is the one where the three-dimensional theory is a boundary of a four-dimensional one. The bulk theory is a $\mathcal{N}=4$ super Yang-Mills $SU(N)$ theory.   The conservation of the stress-tensor fails, due to its non-zero component along the extra dimension. Indeed, in this case there is no shortening condition and the stress-tensor acquires an anomalous dimension, which corresponds to a mass for the dual graviton. \footnote{A similar physical setting is the one where the initial three-dimensional theory is a defect of a four-dimensional one and accordingly we have two contributions of anomalous dimensions, from the left and the right of the defect.} Finally an important detail is that since we are interested in a slightly massive graviton, and therefore in a small anomalous dimension, then it is important that the stress tensor should dissipate \textit{weakly} in the extra dimension. Weak dissipation is ensured by the scarcity of the degrees of freedom of the four-dimensional theory in comparison to the three dimensional boundary one, or by the fact that the boundary degrees of freedom couple weakly to the bulk theory.

\subsection{Extended six-manifolds}
The objective is to formulate solutions with multiple massive gravitons. In previous works \cite{Kiritsis:2006hy}\cite{Kiritsis:2008at} such solutions have been studied as duals to multiple large-N CFTs coupled via mutitrace deformations, in contrast with the common global symmetry gauging that characterises the mechanism in the case at hand. The starting point is (as in the single massive graviton case) the factorisation of the singularities on the strip. We can start by further separating the strip in blocks of five-brane singularities and continue by connecting them gradually. 
\hfill\break
Separating blocks (namely sets of $(N_{i},\hat{N}_{i})$ D5 and NS5 branes) of point-singularities on the two strip boundaries by infinite distance $\delta_{i,i+1}\rightarrow\infty$ initially gives a collection of isolated vacua, namely a collection of non-interacting massless gravitons, the wave functions of which are constant on each spacetime.  Then we proceed with the fist two vacua, by making the distance of the first two blocks of singularities large but finite. The mixing of the first two two massless gravitons, $\psi_{(1)}$ and $\psi_{(2)}$ results to one massless $\Psi_{(1\vert 2)}$ constant over $\mathcal{M}_{6}^{(1)}\cup\mathcal{M}_{6}^{(2)}$ and one massive ($\tilde{\Psi}_{(1\vert 2)}$) combination, interpolating between constant values on $\mathcal{M}_{6}^{(1)}$ and $\mathcal{M}_{6}^{(2)}$. Proceeding to the next vacuum, the initially massless mode of the $\mathcal{M}_{6}^{(3)}$, $\psi_{(3)}$ mixes with the massless mode, $\Psi_{(1\vert 2)}$. This is of course an assumption which can only be made in the case where each throat is very long and therefore in this approximation the wavefunction of the new region mixes only with the one that is constant everywhere in the geometry and not with the one of the first massive graviton. 
\hfill\break
This results to a new massless mode, $\Psi_{(1\vert 3)}$ and a massive mode $\tilde{\Psi}_{(2\vert 3)}$. The final low-lying spectrum therefore includes: one massless mode with a constant wavefunction $\Psi_{(1\vert 3)}$ and two massive: one with wavefunction $\tilde{\Psi}_{(1\vert 2)}$ which varies inside the first Janus throat (connecting regions $\mathcal{M}_{6}^{(1)}$ and $\mathcal{M}_{6}^{(2)}$) 
and one with wavefunction $\tilde{\Psi}_{(2\vert 3)}$ which varies inside the second Janus throat (connecting regions $\mathcal{M}_{6}^{(2)}$ and $\mathcal{M}_{6}^{(3)}$) . This can be generalised to a geometry with N-regions connected by (N-1)-throats, where the spectrum include one massless and (N-1) massive gravitons.

\begin{center}
\begin{figure}[ht]
\centering
\includegraphics[width=.7\textwidth,scale=0.7,clip=true]{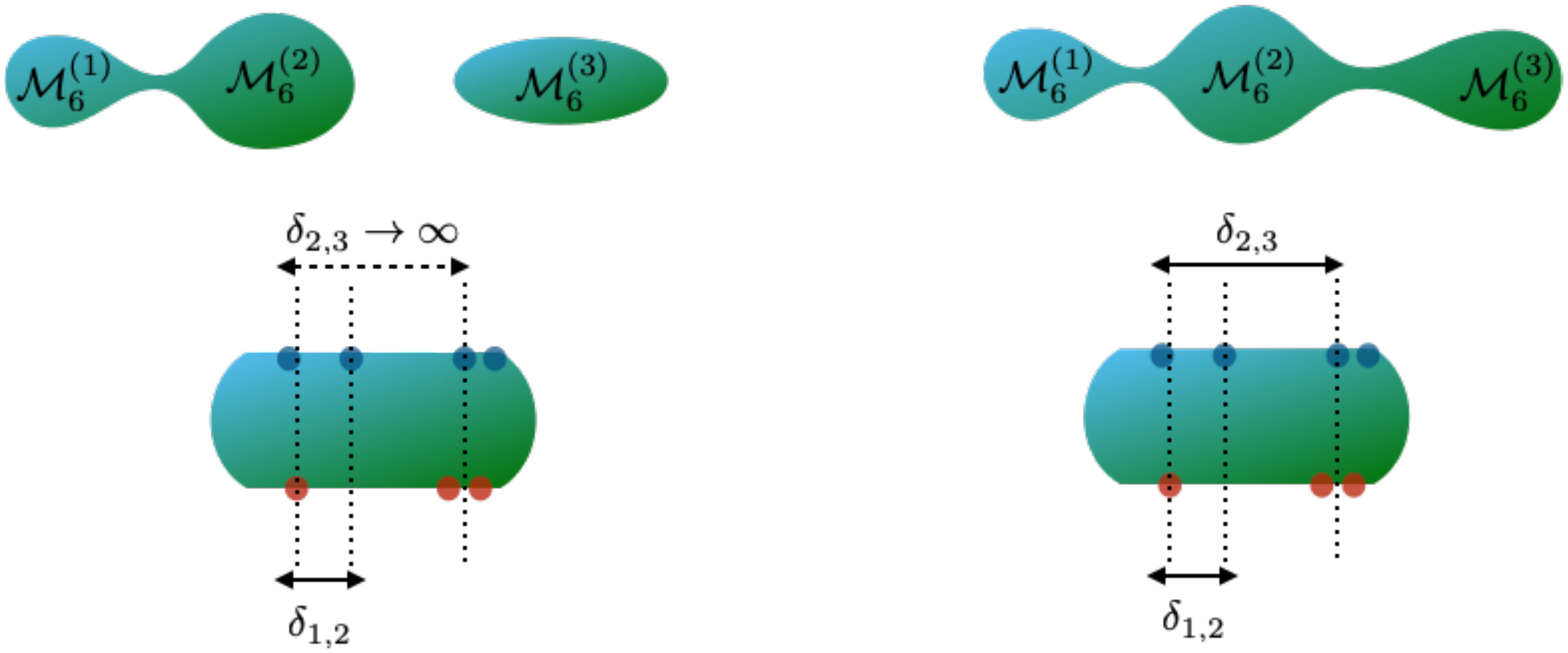}
\caption{Connecting isolated vacua.}
\end{figure}
\end{center}

By Taylor-expanding around the singularities the harmonic functions which parametrise the supergravity solution on the strip we can calculate the geometry of each throat region:

\begin{equation}
h=-\sum\limits_{i=1}^{N}\gamma_{i}\text{logtanh}\Big(\frac{-(x-\delta_{i})+i(\pi/2-y)}{2}\Big)+\text{c.c}=4e^{-\vert x-\delta_{i}\vert}\text{sin}(y)-\frac{4}{3}e^{-\vert x-\delta_{i}\vert}\text{sin}(3y)+\mathcal{O}(e^{-5\vert x-\delta_{i}\vert})
\end{equation} 
\begin{equation}
\hat{h}=-\sum\limits_{\hat{i}=1}^{\hat{N}}\hat{\gamma}_{\hat{i}}\text{logtanh}\Big(\frac{(x-\hat{\delta}_{\hat{i}})+iy}{2}\Big)+\text{c.c}=4e^{-\vert x-\delta_{i}\vert}\text{cos}(y)+\frac{4}{3}e^{-\vert x-\hat{\delta}_{\hat{i}}\vert}\text{cos}(3y)+\mathcal{O}(e^{-5\vert x-\hat{\delta}_{\hat{i}}\vert})
\end{equation}
\hfill\break
, where $\delta_{i},\hat{\delta}_{i}$ are the positions of the D5 and NS5 branes respectively on the two strip boundaries, while $\gamma_{i},\hat{\gamma}_{i}$ stands for the number of each type of five-brane per point singularity. For calculating the geometry of the throat  we note that from the above expansions the leading order terms are kept away from five-brane singularities, while the next-to-leading ones should be kept when considering regions close to 5-brane singularities. The higher order terms are omitted in the above expansions. Everywhere else the leading and next-to-leading order terms suffice. In particular, the latter ones are needed in order for the solution not be singular.  The computation of the geometry of the throat involves only the leading order contributions from the both left and right extremal region five branes, by omitting the next-to-leading order terms. The calculation then results to an $AdS_{5}\times S^{5}$ geometry. If dilaton variation is turned on in the expressions of the harmonic functions (see Appendix) then the throat has Janus geometry.
\hfill\break
Therefore the most general final configuration, is comprised by N-spacetimes connected by (Janus) throats with different quantised charges. The remaining part is to solve the respective minimisation problem for each throat, write down the expression of the corresponding wavefunction and the mass eigenvalue. By following almost the same procedure as the one for the case of the single massive graviton, we obtain the following: First, the normalised wavefunction of the massless mode reads:
\hfill\break
\begin{equation}
\tilde{\Psi}_{1\vert N}\sim \big(\sum\limits_{i=1}^{N}u_{i}\Big)^{-1/2};\quad u_{i}=\mathcal{V}_{6}^{(i)}\langle L_{4}^{2}\rangle_{(i)}
\end{equation}
\hfill\break
,where $u_{i}$ are expressed in terms of the initial constant wavefunctions of the $AdS_{4}\times_{w}\mathcal{M}_{6}$ spacetimes before their mixing and are calculated using the inner products:
\hfill\break
\begin{equation}
\langle \psi_{i}\vert \psi_{i}\rangle=1\rightarrow \psi_{i}^{-2}=\int\sqrt{g}L_{4}^{2}=\mathcal{V}_{6}\langle L_{4}^{2}\rangle_{\mathcal{M}_{i}}\equiv u_{i}
\end{equation}
\hfill\break
The minimisation problem reads:

\begin{equation}
m_{0}^{2}\sim\text{min}_{\psi}\Big(\int d^{6}y\sqrt{g}L_{4}^{4}g^{ab}\partial_{a}\tilde{\Psi}_{(j\vert j+1)}\partial_{b}\tilde{\Psi}_{(j\vert j+1)}\Big)\sim\int\limits_{x_{j}^{+}}^{x_{j+1}^{-}}dx L_{5(j\vert j+1)}^{8}\mathcal{G}(x)\Big(\frac{d\Psi_{(j\vert j+1)}}{dx}\Big)^{2}
\end{equation}
\hfill\break
where $\mathcal{G}(x)=\Big(\frac{\text{cosh}(\delta\varphi)+\text{cosh}(2x)}{\text{cosh}(\delta\varphi)}\Big)^{2}$. The boundary conditions for each massive-mode wavefunction of the overall configuration read:
\begin{equation}
\Psi_{(j\vert j+1)}\rightarrow
\left\{
	\begin{array}{ll}
		\Psi_{j}  & x\rightarrow x_{j}^{+} \\
		\Psi_{j+1}& x\rightarrow x_{j+1}^{-}
	\end{array}
\right.
\end{equation}
\hfill\break
where $\Psi_{j},\Psi_{j+1}$ are the constant values between which the massive-mode wavefunction is interpolating. These are calculated by using the inner products (all wavefunctions are normalised and the ones corresponding to the massive modes are orthogonal to the one of the massless mode):
\hfill\break
\begin{equation}
\langle\tilde{\Psi}_{(j\vert j+1)}\vert\tilde{\Psi}_{(j\vert j+1)}\rangle=1,\quad \langle\tilde{\Psi}_{(j\vert j+1)}\vert\Psi_{(1\vert j+1)}\rangle=0
\end{equation}
\hfill\break
\begin{center}
\begin{figure}[ht]
\centering
\includegraphics[width=.6\textwidth,scale=0.6,clip=true]{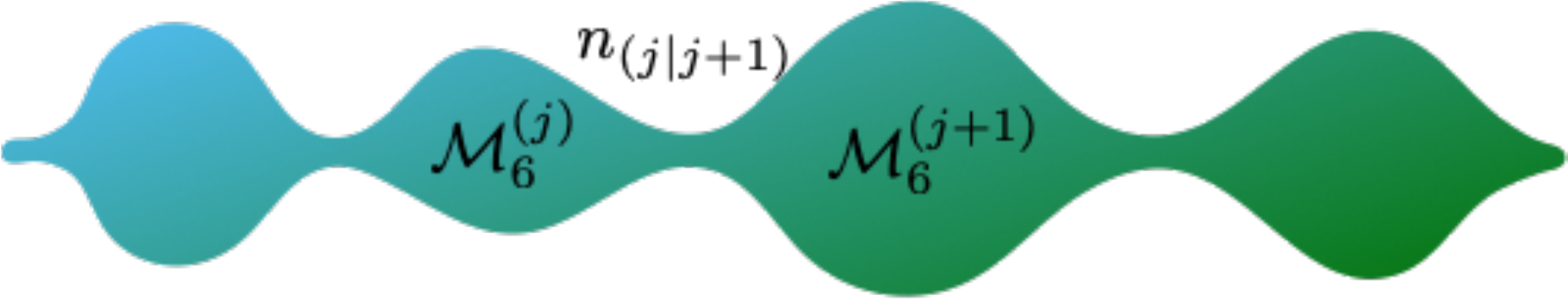}
\caption{The final configuration, including one massless and N-1 massive modes. In the above figure, $n_{j\vert j+1}$ stands for the quantised charge of the Janus throat connecting $\mathcal{M}_{6}^{(j)}$ and $\mathcal{M}_{6}^{(j+1)}$.}
\end{figure}
\end{center}
\hfill\break
From the minimisation problem, we obtain the expression for the wavefuntion of a massive graviton. The constant values between which the wavefunction interpolates are given below:
\hfill\break
\begin{equation}
\tilde{\Psi}_{(j\vert j+1)}\sim
\left\{
	\begin{array}{ll}
		(u_{j+1}u_{j}^{-1})^{1/2}(u_{j+1}+u_{j})^{-1/2} & \mbox{if } x< x_{j^{+}} \\
				-(u_{j+1}^{-1}u_{j})^{1/2}(u_{j+1}+u_{j})^{-1/2} & \mbox{if} x>x_{j+1^{-}}
	\end{array}
\right.
\end{equation}
\hfill\break
, while in the throat region, we obtain:
\hfill\break
\begin{equation}
\tilde{\Psi}_{(j\vert(j+1))}(x)\sim\frac{1}{2(u_{j}u_{j+1}(u_{j}+u_{j+1}))^{1/2}}\Big((u_{j+1}-u_{j})-(u_{j+1}+u_{j})\frac{I(x,\alpha)}{I(x_{(j+1)},\alpha)}\Big)
\end{equation}
\hfill\break
where $I(x,\alpha)=\int\limits_{0}^{x}dw\frac{\alpha^{2}}{(cosh(w)+\alpha)^{2}}$ and $\alpha=cosh(\delta\varphi)$.
\hfill\break
\hfill\break
The leading order mass in this long Janus throat approximation is then obtained by substituting the above expression in the eigenvalue equation:
\hfill\break
\begin{equation}
m_{j}^{2}L_{4}^{2}\simeq \frac{3n_{(j-1\vert j)}^{2}}{16\pi^{2}}\Big(\frac{\kappa_{4(j-1)}^{2}}{\langle L_{4}^{2}\rangle_{(j-1)}}+\frac{\kappa_{4(j)}^{2}}{\langle L_{4}^{2}\rangle_{(j)}}\Big)\times\mathcal{J}(cosh(\delta\varphi_{(j-1\vert j)}))
\end{equation}
\hfill\break
\hfill\break

\begin{flushleft}
\textbf{Circular geometry}
\end{flushleft}

We close this section with an interesting case, similar to the one described above. It is the one where the two six-manifolds are connected by \textit{two} throats. This framework is valid, as the considered solutions are such that a six-manifold can have (at most) two throats attached on it. This configuration can be obtained as a dual to a circular quiver \cite{Assel:2012cj}, where two of the unitary groups have low rank. Then our geometry results from a double pinching of the initial annulus (which is the geometry of the supergravity solutions dual to circular quivers). A factorisation of the singularities of the above compact six-manifold leads to the configuration of two bags connected by two thin throats. The spectrum now includes again one massless and one massive graviton, the mass of which in this case depends on the charges of the two throats.

\begin{center}
\begin{figure}[ht]
\centering
\includegraphics[width=.6\textwidth,scale=0.6,clip=true]{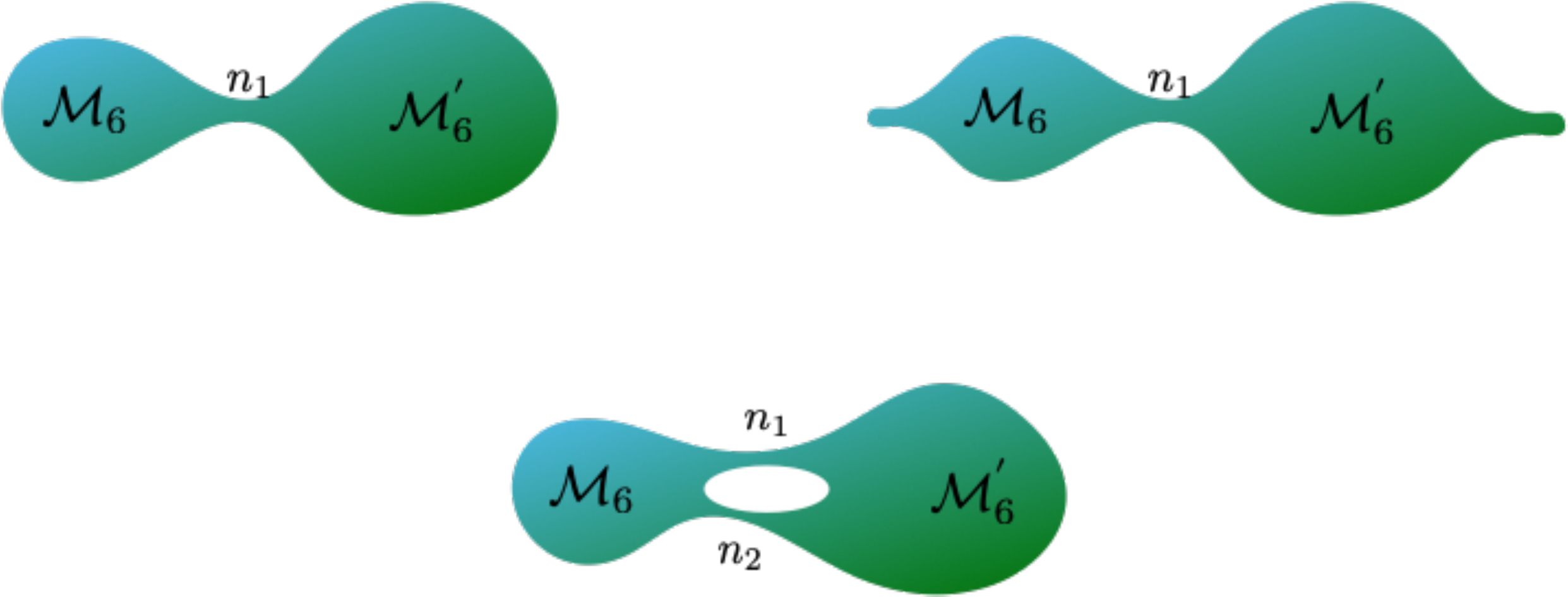}
\caption{Connection of two compact six-manifolds by two highly curved Janus throats, as explained in the main text.}
\end{figure}
\end{center}
Indeed, the massive graviton wave function is a combination of the initial constant wavefunctions which vanishes inside both throats while interpolating between constant values in the two spacetimes. The leading order mass is given below. Note that this time it is suppressed by the square of the Janus correction function $\mathcal{J}$, as dilaton variations along both throats are taken into account:

\hfill\break
\begin{equation}
m_{g}^{2}L_{4}^{2}\sim \frac{3(n_{1}^{2}+n_{2}^{2})}{16\pi^{2}}\Big(\frac{\kappa_{4(1)}^{2}}{\langle L_{4}^{2}\rangle_{(1)}}+ \frac{\kappa_{4(2)}^{2}}{\langle L_{4}^{2}\rangle_{(2)}}\Big) \times \mathcal{J}(cosh(\delta\varphi))^{2}
\end{equation}

\hfill\break

The dual picture can also be seen as two 3d defects in two distinct 4d bulks:
The two three-dimensional theories are localised on the 3d boundaries of two separate four-dimensional bulks, where $\mathcal{N}=4$ 4d SYM theories with gauge groups $SU(n_{1})$, $SU(n_{2})$ and couplings $g_{YM}^{(1)}$, $g_{YM}^{(2)}$ live (Figure 6.)

\begin{center}
\begin{figure}[ht]
\centering
\includegraphics[width=.52\textwidth,scale=0.6,clip=true]{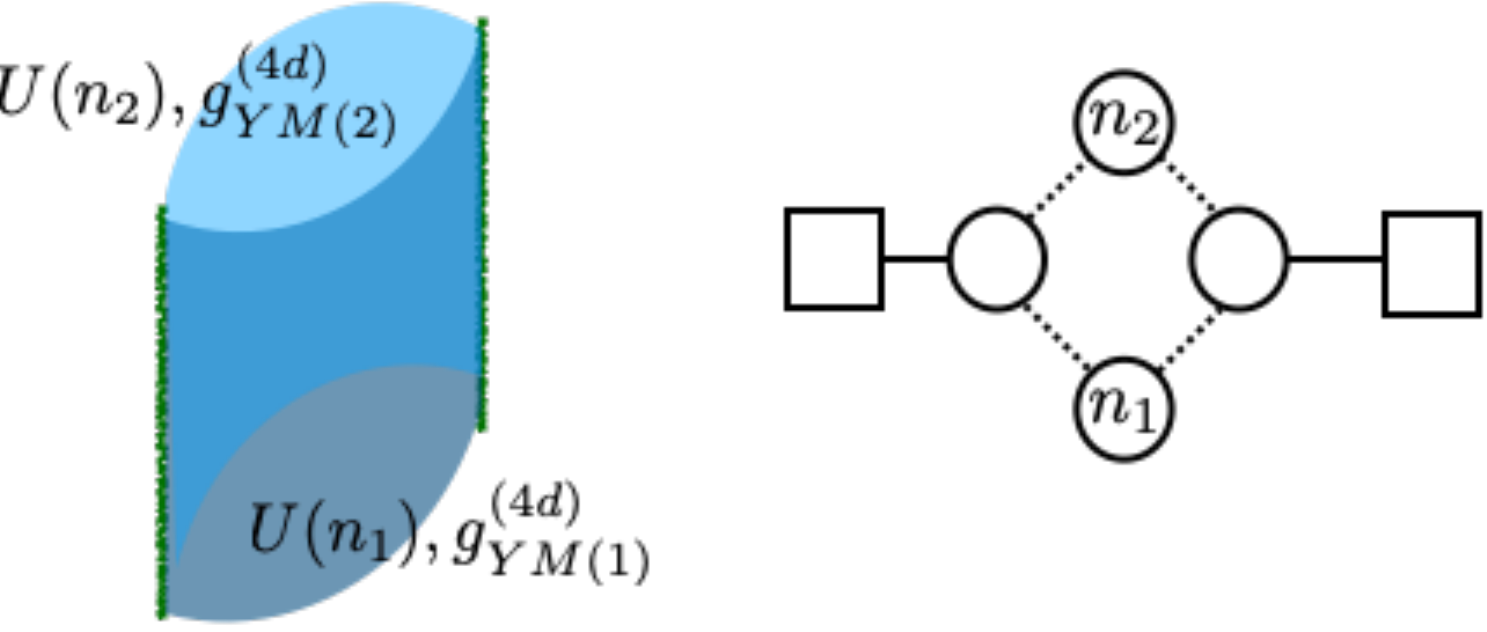}
\caption{The dual picture of the configuration where two $AdS_{4}\times_{w}\mathcal{M}_{6}$ spacetimes are connected by two Janus throats.}
\end{figure}
\end{center}

%%%%%%%%%%%%%%%%%%%%%%%%%%%%%%%%%%%%%%%%%%%%%%%%%%%%%%%%%%%%%%%%%%%%%%%%%%%%%%%%%%%%%%%%%%%%%%%%%%%%%%%%%%%%%

%%%%%%%%%%%%%%%%%%%%%%%%%%%%%%%%%%%%%%%%%%%%%%%%%%%%%%%%%%%%%%%%%%%%%%%%%%%%%%%%%%%%%%%%%%%%%%%%%%%%%%%%%%%%%

 \section{Generalised Distance Conjecture and Massive Gravity}
 
 In this section we are going to study the aforementioned embeddings under the scope of the generalised distance conjecture. The well known swampland criterion of the generalised distance conjecture \cite{Lust:2019zwm} predicts for quantum gravity in d-dimensional Riemannian manifolds a tower of light states appearing at infinite distance in scalar moduli space, with mass scale:
 \hfill\break
 \begin{equation}
 m\sim e^{-\alpha\Delta_{\phi}}\, ,
 \end{equation}
 \hfill\break 
 ,where $\alpha$ is a positive parameter the value of which will be discussed below and $\Delta_{\phi}$ stands for the distance in the space of the modulus $\phi$. It is given by the expression:
 \hfill\break
 \begin{equation}
 \Delta_{\phi}\sim\int\limits_{\tau_{i}}^{\tau_{f}}d\tau\Big(p_{ij}(\phi)\partial_{\tau}\phi^{i}\partial_{\tau}\phi^{j}\Big)^{1/2}
\hfill\break
\end{equation}
In the above expression the label $\tau$ parametrises the path in the moduli space and $p_{ij}(\phi)$ signifies the metric in field space.
\hfill\break
Its generalised version, regards distances for all background fields in the theory. In particular when applied to variations of the metric, the conjecture no longer refers to distance in the moduli space but rather to the distance  $\Delta_{g}$ in the space of background metrics. With the metric of the manifold labelled by a "time" parameter $g_{\mu\nu}(\tau)$, the generalised conjecture predicts an infinite tower of light modes, at infinite distance in the metric space appearing at scale:
 \hfill\break
 \begin{equation}
 m(\tau_{f})\sim m(\tau_{i})e^{-\alpha\vert\Delta_{g}\vert}\, , 
 \end{equation}
 \hfill\break
where the mass at an initial "time" is usually considered to be $m(\tau_{i})\sim M_{\text{Pl}}$, and the distance is given by the expression:
\hfill\break
\begin{equation}
\Delta_{g}\sim\int\limits_{\tau_{i}}^{\tau_{f}}d\tau\Big(\mathcal{V}_{\mathcal{M}}^{-1}\int\sqrt{g}g^{\mu\nu}g^{\rho\sigma}\partial_{\tau}g_{\mu\rho}\partial_{taug_{\nu\sigma}}\Big)^{1/2}\, ,
\end{equation}
\hfill\break
with $\mathcal{V}_{\mathcal{M}}=\int_{\mathcal{M}}\sqrt{g}$ being the volume of the d-dimensional manifold. 
\hfill\break
In the formulation of the generalised distance conjecture for supersymmetric AdS solutions of string/M-theory, the infinite tower appears as we approach $\Lambda\rightarrow 0$, and scales as:
\hfill\break
\begin{equation}
m\sim\vert\Lambda\vert^{1/2}\, ,
\end{equation}
\hfill\break
This constitutes the strong version of the distance conjecture on AdS space, where $\alpha=\frac{1}{2}$, which is verified for a number of solutions and has been correlated, apart from the original distance conjecture, to other conjectures of the Swampland program, such as the refined distance conjecture for de-Sitter space \cite{Ooguri:2018wrx}. This is also correlated with the absence of scale separation of AdS flux vacua \cite{Tsimpis:2012tu,Gautason:2015tig,Gautason:2018gln}: In above limit, where practically the inverse AdS radius vanishes, the volume of the internal manifold- in the case of a background of the type $AdS_{p}\times \mathcal M$- blows up, indicating that the radii $L_{AdS}$ and $L_{\mathcal{M}}$ are not separated but are rather parametrically bound.
\hfill\break
Moreover, a particularly significant observation \cite{Kehagias:2019akr} is that the generalised distance conjecture, when applied for metric variations, is related to geometric flow equations in General Relativity. The Ricci flow equation, introduced by Hamilton \cite{hamilton1982} , is the evolution equation that describes the deformation of the metric $g_{\mu\nu}(\tau)$ on a Riemannian d-dimensional manifold $\mathcal{M}_{d}$, driven by its curvature:
\hfill\break
\begin{equation}
\partial_{\tau}g_{\mu\nu}(\tau)=-2R_{\mu\nu}(\tau)\, .
\end{equation}
\hfill\break
 Fixed points of the Ricci flow, where $\mathcal{R}=0$ are found at infinite distance in metric space where the infinite tower is conjectured to appear and correspond to solutions of string theory. For the case of string backgrounds which include also scalar fields (as the dilaton field in our case) in place of just the Ricci flow, a refined set of evolution equations now gives a well defined distance measure. The distance is now given in terms of the so called Entropy Functional, introduced by Perelman \cite{perelman2002entropy}
 \hfill\break
 \begin{equation}
 \mathcal{F}(g,f):=\int d^{d}x\sqrt{-g}(\mathcal{R}+(\nabla f)^{2})e^{-f}\, ,
 \end{equation}
 \hfill\break
describing that the Ricci flow can be formulated as a gradient flow. Considering variation of the functional with respect to the metric and to the scalar field, while preserving the volume form $\int d^{d}x\sqrt{-g}e^{-f}$, leads to the refined set of equations:
\hfill\break
\begin{align*}
&\partial_{\tau}g_{\mu\nu}(\tau)=-2R_{\mu\nu}(\tau)\\
&\partial_{\tau}f(\tau)=-\mathcal{R}(\tau)-\nabla^{2}f(\tau)+(\nabla f(\tau))\, ,
\end{align*}
\hfill\break
Along these lines, the distance conjecture states that the distance along the Ricci-dilaton flow is given by $\Delta_{\mathcal{F}}\sim \text{log}\mathcal{F}$, while at the fixed point of the combined flow, where $\mathcal{F}=0$, there is an infinite tower of massless modes.
\hfill\break

\subsection{Tower of states and Effective massive-gravity theory}

The above brief introduction to the generalised distance conjectures in quantum gravity and to the notion of geometric flows, will now be put in the context of bimetric and massive gravity on $AdS_{4}$. As described in the previous section, the main (leading order) contribution to the mass of the lowest lying spin 2 mode comes from the Janus throat, with metric:
\hfill\break
\begin{equation}
ds^{2}_{\text{Janus}}=l_{5}^{2}dx^{2}+l_{4}(x)ds^{2}_{AdS_{4}}\, ,
\end{equation}
\hfill\break
,where $l_{4}(x)$ is the radius of the $AdS_{4}$ fiber. The throat is capped-off on the left and the right by $AdS_{4}\times_{w}\mathcal{M}_{6}^{(1,2)}$ for the case of bigravity, while capped off only on the one side and approximating  $AdS_{5}\mathbb{Z}_{2}\times S^{5}$ for the case of massive gravity.  In the $\delta x=x_{R}-x_{L}$ interval (throat) $l_{4}(x)\sim l_{5}$.
\hfill\break
According to the statement of the recent massive graviton distance conjecture \cite{Bachas:2019rfq}, at infinite distance in scalar moduli space, $\delta\varphi\rightarrow\infty$ , the Janus  throat develops a flat region, where $m_{g}\rightarrow 0$ and an infinite tower of KK modes appears.  The approach to the question of the breakdown scale of the effective field theory is particularly interesting. According to the universal statement for $AdS_{4}$, obtained by the analysis of the non-linear action for the Stückelberg field \cite{deRham:2016plk}, the (highest) scale at which strong coupling sets in, is :
\hfill\break
\begin{equation}
\Lambda_{*}=\Big(\frac{m_{g}M_{\text{Pl}}}{L_{4}}\Big)^{1/3}\, ,
\end{equation}
\hfill\break
as can be seen from the interaction terms between the (normalised) scalar fields in the action. In terms of the dual three-dimensional CFT, as shown in \cite{Bachas:2019rfq} , this scale corresponds to a bound-scaling dimension $\Delta_{*}=3+(\varepsilon_{g}c)^{1/3}$, where $\varepsilon_{g}c<<1$ and is the scale at which new spin$\geq2$-operators enter the spectrum. This is dual to the tower of spin$\geq2$ modes at scale $\Lambda_{*}$, which is origin of the breakdown of the effective field theory \footnote{Note that in \cite{Bachas:2019rfq}, it is explained that KK modes actually start descending at scales lower than $\Lambda_{*}$}. According to the respective massive-AdS-graviton distance conjecture, this tower descends as we approach the point of scalar moduli space where $m_{g}\rightarrow 0$, which lies at infinite distance. For the multiple graviton solutions, the strong coupling scale should be determined by the highest mass eigenvalue in the configuration.
\hfill\break
\hfill\break
As can be seen by what was described in the beginning of the section, the above statement is in agreement with the arguments of the distance conjectures. In the solution at hand, the only modulus is the dilaton field, for which its is straightforward to see that $\Delta\sim\delta\varphi$. For infinite dilaton variation, namely for infinite distance in moduli space, the conjecture predicts the an infinite tower appearing at a flat region of the manifold. The tower includes massless modes (KK modes of the internal manifold, for the masses of which we have $m_{KK}\sim\frac{1}{l_{4}}\rightarrow 0$ and finally string modes and/or multi-particle bound states. This region should in principle be the one where $m_{g}\rightarrow 0$ in the case at hand and where the effective field theory should break down, with the breaking originating from the appearance of the tower.

We can in principle try to compare the two descriptions in this level. Motivated by the above results, one can expect that the scales $m$ and $\Lambda_{*}$ can be related in this particular context.
\hfill\break 
 For the simple case of massive gravity where the geometry of the throat is $AdS_{5}\times S^{5}$, the mass of the lowest lying graviton is given by the following expression:
\hfill\break
\begin{equation}
m^{2}_{g}L_{4}^{2}\sim\frac{3n^{2}}{16\pi^{2}}\frac{\kappa_{4}^{2}}{\langle L_{4}^{2}\rangle_{\mathcal{M}_{6}}}
\end{equation}
\hfill\break
and the scale of the tower in the limit where $\Lambda\rightarrow0$ is given as\, .
\begin{equation}
m\sim \vert\Lambda\vert^{\alpha}
\end{equation}
The value of the $\alpha$-parameter for supersymmetric $AdS$-vacua is bounded $\alpha\geq\frac{1}{2}$. Here we follow the strong version of the conjecture according to which for supersymmetric AdS vacua $\alpha=\frac{1}{2}$; it will become obvious from power counting that this is the only allowed value in the context at hand.
Substituting $\alpha$ and combining the two expressions for the scales we obtain:
\begin{equation}
 m\sim\frac{m_{g}^{\frac{1}{2}}}{(\kappa_{4}n)^{\frac{1}{2}}}=\frac{M_{\text{Pl}}^{\frac{1}{2}}m_{g}^{\frac{1}{2}}}{n^{\frac{1}{2}}}\, .
\end{equation}
 By substituting now the radius $L_{4}$ in the $\Lambda_{*}$ expression, we obtain:
\begin{equation}
\Lambda_{*}\sim\frac{M_{\text{Pl}}^{\frac{1}{2}}m_{g}^{\frac{1}{2}}}{n^{\frac{1}{6}}}\, .
\end{equation}
The two scales can now be compared and the result reads:
\begin{equation}
\frac{m}{\Lambda_{*}}\sim\frac{1}{n^{1/3}}\, .
\end{equation}
 The result indicates that the two scales are of the same order (and in particular coincide for minimal five-form flux in the throat $n=1$). This agrees with the initial assumption and shows that the (generalised) distance conjecture is verified in the case of massive gravity on $AdS_{4}$. The above description, indicates that there is a threefold-approach to the breakdown scale: On the one hand, there is the universal result of \cite{deRham:2016plk} for $\Lambda_{*}$ obtained directly from the action for the Stückelberg field. The origin of the breakdown is explained holographically in \cite{Bachas:2019rfq} and is traced to the appearance of a tower of states at scale $\Lambda_{*}$, Finally, the analysis in the context of the generalised infinite distance conjecture,  results to a tower of states appearing at scale $m\sim\Lambda_{*}$, indicating the breakdown scale of the effective theory.
\hfill\break
We close this section with a notable remark which regards the expression of the scale of the infinite tower. The generalised distance conjecture is formulated for  quantum gravity on d-dimensional Riemannian manifolds, independent of modifications as non-vanishing mass for the graviton. In the case at hand, we have an explicit a formula for the graviton mass with respect to the geometric/gravitational parameters of the configuration. Therefore in this particular example of massive gravity, the result for the infinite tower scale as given by this  analysis, can be directly re-expressed in terms of the graviton mass, as given above. 

\section{Concluding Remarks}
This threefold approach to the effective field theory breakdown scale of massive and/or multi-metric gravity on $AdS_{4}$, could in principle be realised also in other massive gravity contexts. An interesting problem would be the verification of the compatibility of this approach to other $AdS$-massive gravity embeddings. Some of the candidate $AdS$ supergravities are for $d=5, \mathcal{N}=2$, which is dual  to the well known Class-$\mathcal{S}$ SCFTs \cite{Gaiotto:2009we,Gaiotto:2009gz,Distler:2017xba} and for $d=4,\mathcal{N}<3$. The common feature of these supergravities is that the stress-tensor of the dual SCFTs is not protected and therefore can acquire anomalous dimension, which is dual to small non-zero mass for the graviton. One should be able to calculate the mass of the lowest-lying spin 2 mode (and also study the extension to multiple interacting gravitons), study the origin of the breakdown  of the effective theory and verify that the same result is obtained in the context of the generalised distance conjecture, reinforcing in this way its validity.
\hfill\break
Finally, another interesting question regards to the Minkowski limit of the $AdS_{4}$ (multi)gravity. The limit $L_{4}^{-1}\rightarrow 0$ takes us from $AdS_{4}$ where the breakdown scale is $\Lambda_{*}$, to $\mathbb{M}_{4}$ with the breakdown scale being $\Lambda_{\mathbb{M}_{4},*}=(m^{2}M_{\text{Pl}})^{1/3}$. It would be interesting to study this limit under the above scope and moreover make contact with the proposed spin-2 conjectures of \cite{Klaewer:2018yxi}.

\hfill\break
\hfill\break
\textbf{Acknowledgements:}
\hfill\break
We thank Costas Bachas, Dongsheng Ge and Eran Palti for useful discussions and for suggestions on the manuscript. The work of D.L and I.L is supported by the Origins Excellence Cluster.

\begin{flushleft}
\Huge \textbf{Appendix}
\end{flushleft}
\appendix
\addcontentsline{toc}{section}{Appendix}

\input{Appdx.tex}

\newpage

\bibliography{LLbibj}{}
\bibliographystyle{unsrt}

\end{document}

%% file: Appdx.tex
\begin{flushleft}
\textbf{A: IIB supergravity solutions}
\end{flushleft}
The  solutions of type-IIB supergravity that preserve ${\cal N}=4$ AdS$_4$
    supersymmetries have the  general  form   of an  AdS$_4$ and 
     two 2-sphere   fibers warped  over a Riemann-surface, 
     \hfill\break
     \bea ds^2_{10} \, =\,    L_4^{\,2}   ds^2_{{\rm AdS}_4}  +                                                                                                                                                                                                                                                                                                                                                                                                                                                                                                                                                                                                                                                                                                                                                                                                                                   f^{\,2} ds^2_{{\rm S}^2} + \hat{f}^{\,2} ds^2_{{\rm \hat{S}}^2} + 4\rho^2 dz d\bar z  
  \eea  
  \hfill\break
    where all scale factors  depend only on  the coordinates $(z,\bar z)$ of the Riemann surface which in our case is
      the  infinite strip
    $0\leq {\rm Im}z \leq \pi/2$. The  expressions for the metric factors and for the dilaton $\phi$ 
    depend on  two harmonic functions $h_1, h_2$
    \cite{D'Hoker:2007xy}\cite{D'Hoker:2007xz}:
    \hfill\break
    \bea\label{17}
    L_4^{\,8}= 16\, {{ \cal U} {\hat{\cal U}}\over  W^2} \ ,   \quad 
 f^{\,8} = 16 \,\hat{h}^{\,8} \,{ {\hat{\cal U}}W^2 \over {\cal U}^{\,3}} \ ,
 \quad
\hat{f}^{\,8} = 16 \,h^{\,8}\, { {\cal U} W^2 \over {\hat{\cal U}}^{\,3}} \ , \quad 
  \rho^8 = {{ \cal U}  {\hat{\cal U}} W^2 \over h^4\hat{h}^4} \ ,  
  \quad e^{4\phi} = { {\hat{\cal U}}   \over {\cal U}}\  ,  
 \eea
where ,
\hfill\break
\hfill\break
   \bea\label{a1}
\ { \cal U} = 2 h\hat{h}|\partial_z h\vert^2 - h^2 W\  \qquad \ {\hat{\cal U}} = 2 h\hat{h}|\partial_z \hat{h}\vert^2 - \hat{h}^2 W\  {\rm and} \qquad 
W = \partial_z \partial_{\bar{z}} (h\hat{h}) \ .
   \eea

  %%%%%%%%%%%%%%%%%%%%%%%%%%%%%%%%%%%%%
  The solutions include non-vanishing R-R and NS-NS three-form field strengths:

\hfill\break
\begin{equation}
F_{(3)}=\hat{\omega}_{(2)}\wedge db_{2}, \quad H_{(3)}=\omega_{(2)}\wedge db_{1}
\end{equation}
\hfill\break

where $\hat{\omega}_{(2)}$, $\omega_{(2)}$ stand for the volume forms of $\hat{S}^{2},S^{2}$ and the $b_{1,2}$ are one forms which depend on the harmonic functions. The remaining components of the supergravity solution are the R-R 5-form and the dilaton:

\hfill\break
\begin{equation}
F_{5}=-4L_{4}^{4}\omega_{(4)}\wedge\mathcal{F}+4f^{2}\hat{f}^{2}\omega_{(2)}\wedge\hat{\omega}_{(2)}\wedge(\star_{(2)}\mathcal{F}),\qquad e^{4\phi}=\frac{\hat{\mathcal{U}}}{\mathcal{U}}
\end{equation}
\hfill\break

,with $\omega_{(4)}$ being the $AdS_{4}$-volume form and $\mathcal{F}$ being a 1-form on the infinite strip with the property that the quantity $L_{4}^{4}\mathcal{F}$ is closed. 
It is straightforward to prove the following identities that hold between the metric factors and the dilaton: $L_{4}^{2}f^{8}=4e^{2\phi}h^{2}$,  $L_{4}^{2}\hat{f}^{8}=4e^{-2\phi}\hat{h}^{2}$ and $\rho^{6}f^{2}\hat{f}^{2}=4W^{2}$. 
\hfill \break
\hfill \break
A trait of the solutions is the fact that the harmonic functions $h,\hat{h}$ obey particular boundary conditions on the two strip-boundaries. The first harmonic function vanishes on the lower boundary along with the normal derivative of the second harmonic function while the converse holds for the upper boundary. In this sense the $\Sigma_{(2)}$-boundary is realized in terms of divisions:

\hfill\break
\begin{equation}
h\vert_{y=0}=\partial_{\perp}\hat{h}\vert_{y=0}=0 \quad \text{and} \quad \hat{h}\vert_{y=\frac{\pi}{2}}=\partial_{\perp}h\vert_{y=\frac{\pi}{2}}=0
\end{equation}
\hfill\break

The above conditions imply that the $f$-factor vanishes in the lower strip-boundary whereas the $\hat{f}$-factor vanishes on the upper strip-boundary. This, in turn, corresponds to the fact that the radius of the $S^{2}$ vanishes on the lower strip boundary and accordingly the radius of the $\hat{S}^{2}$ vanishes on the upper boundary. Therefore, the points on the $\Sigma_{(2)}$-boundary are interior points of the full ten-dimensional geometry; the only boundary in this framework is the conformal boundary of $AdS_{4}$.

These conditions, along with the requirement that both the dilaton and the metric factors be regular in the interior of $\Sigma_{(2)}$ but divergent at isolated points of the boundaries, form a set of global consistency conditions that severely constrain the supergravity solutions and directly the form of the harmonic functions $(h,\hat{h})$. Regarding the aforementioned isolated points of the $\Sigma_{(2)}$ boundary, there are three kinds of admissible singularities for the harmonic functions, interpreted to be sourced by type IIB branes: D5, NS5 and D3 branes. The ones interpreted as fivebranes, are logarithmic-cut singularities and in what follows we will see that they also carry D3 brane charge. As regards the ones interpreted purely as D3 brane sources, they are square-root singularities\cite{DHoker:2007zhm}. 

We can proceed to the detailed description of the simplest solution which contains brane singularities. The form of the two harmonic functions is given below:

\hfill\break
\begin{align}
&h=-i\alpha\text{sinh}(z-\beta)-\underline{\gamma\text{log}\Big[\text{tanh}\big(\frac{i\pi}{4}-\frac{z-\delta}{2}\big)\Big]}+\text{c.c} \\
&\hat{h}=\hat{\alpha}\text{cosh}(z-\hat{\beta})-\underline{\hat{\gamma}\text{log}\Big[\text{tanh}\big(\frac{z-\hat{\delta}}{2}\big)\Big]}+\text{c.c}
\end{align}
\hfill\break

  %%%%%%%%%%%%%%%%%%%%%%%%%%%%%%%%%%%%%%%
   
All the parameters in these expressions are  real. Furthermore the  
 $\{\alpha, \gamma_a \}$  and $\{\hat{\alpha},  \hat\gamma_b\}$ must have the same sign 
 which 
can be  chosen positive so that $h, \hat{h} >0$   in the interior of  the strip.  
The logarithmic singularities on the upper (lower) boundary  of the strip  correspond to    D5-brane (NS5-brane)
sources wrapping closed  submanifolds of M$_6$. The regions $z\to\pm\infty$  correspond to two  semi-infinite
D3-brane throats whose geometry is that of the  supersymmetric  Janus.

The two harmonic functions are labelled by two sets of  four real parameters, $(\alpha,\beta,\gamma,\delta)$ and the corresponding hatted ones. The underlined part of the expressions corresponds to the fivebrane singularities: $\delta$ and $\hat{\delta}$ are the positions of the D5 and NS5 sources on the upper and lower strip-boundaries. Obviously this solution describes just one stack of D5 branes on the upper boundary at $z=\delta+\frac{i\pi}{2}$ and one of NS5s on the lower boundary, at $z=\hat{\delta}$.  Except of these singular points, all the rest of the boundary belongs to the interior of the full ten-dimensional geometry. Regarding the rest of the parameters, we will see that $(\beta,\hat{b})$ determine the dilaton and in particular the dilaton variation between the two asymptotic region of the strip. This is explained in detail in the next part of the Appendix. The parameters $(\alpha,\hat{\alpha})$ and $(\gamma,\hat{\gamma})$ play an important role in the physics and the geometry of our setup and are going to be explained below. It is notable that they are assumed to be all non-negative \cite{DHoker:2007zhm}.

Let us now examine the physics in the vicinity of the D5 branes. Recall that the two sphere $\hat{S}^{2}$ shrinks to zero radius on the upper boundary. Consider a segment $I$ surrounding the singularity at $z=\hat{\delta}+\frac{i\pi}{2}$ and the fibration of the $\hat{S}^{2}$ over this segment: $I\times \hat{S}^{2}$ is a non-contractible three-cycle. It has topology of a three-sphere given the fact that on the upper boundary the second harmonic function and the second two-sphere form factor vanish ($\hat{h}=\hat{f}=0$). It can be verified that this three cycle supports a non-vanishing RR three-form flux, which corresponds to the total number of the D5 branes at the particular point (or equivalently to the total D5 brane charge).

Accordingly, focusing on the lower-boundary singularity, we have the non-contractible three-cycle $I'\times S^{2}$obtained as a fibration of $S^{2}$ two-sphere over a segment $I'$, which surrounds the singularity at $z=\hat{\delta}$. This three-cycle (which also has the topology of a three-sphere as on the lower boundary $h=f=0$) supports non-vanishing NS-NS three-form flux, which counts the total number of NS5 branes at that point (or equivalently the total NS5 brane charge).

\hfill\break
\begin{equation}
N_{D5}=\frac{1}{4\pi^{2}\alpha'}\int_{I\times\hat{S}^{2}}F_{(3)}=\frac{4}{\alpha'}\gamma,\quad N_{NS5}=\frac{1}{4\pi^{2}\alpha'}\int_{I'\times S^{2}}H_{(3)}=\frac{4}{\alpha'}\hat{\gamma}
\end{equation}
\hfill\break

, from where the meaning of the real parameters $\gamma,\hat{\gamma}$ becomes evident: they count (in appropriate normalization) the number of fivebranes per stack. Moreover, it should be noted that the fivebrane charge is quantized in units of $2\kappa_{0}T_{5}$, with $2\kappa_{0}^{2}=(2\pi)^{7}\alpha'^{4}$ being the gravitational coupling, and with the fivebrane tension being $T_{5}=\frac{1}{(2\pi)^{5}\alpha'^{3}}$, while since the dilaton is being kept arbitraty, one has the freedom to choose  for the string coupling $g_{s}=1$ which renders the tensions of the two fivebrane types equal.

\hfill\break
\begin{center}
\begin{figure}
\centering
\includegraphics[width=.66\textwidth,scale=0.6,clip=true]{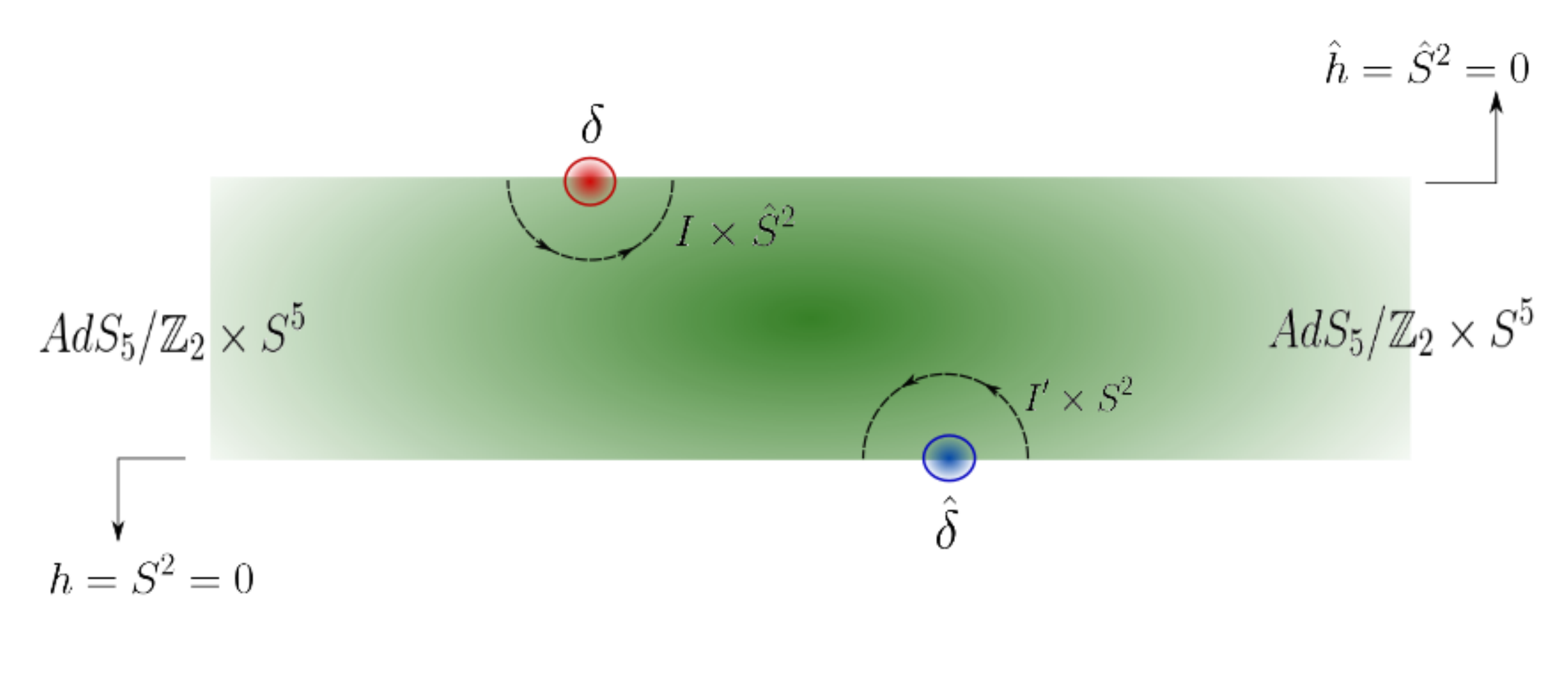}
\caption{Solution with one D5 and one NS5 brane: The indicated  three-cycles at the region of each point singularity, are non-contractible and support R-R and NS-NS three-form fluxes, as explained in the text.  The boundary conditions for the harmonic functions are specified, along with the asymptotic $AdS_{5}/\mathbb{Z}_{2}\times S^{5}$ regions. }
\end{figure}
\end{center}  
\hfill\break

What remains to be presented regarding the singularities is the D3 brane charge of these solutions. First of all, the aforementioned singularities carry D3 brane charge, apart from the fivebrane charge already described. Focusing on the upper-boundary supporting singularities, except from the $F_{3}$ flux supported by the three-cycle $I\times \hat{S}^{2}$, there is also a five-form flux threading the non-contractible five-cycle which is obtained as a fibration of the $S^{2}$ over the above three-cycle:  $(I\times \hat{S}^{2})\times S^{2}$. A significant detail is that this five-form is not just simply the R-R $F_{5}$, but rather the five-form $\tilde{F}_{5}= F_{5}-B_{2}\wedge F_{3}$, with $B_{2}$ being the gauge potential of the NS-NS three-form.  Accordingly, regarding the lower-boundary singularities, except from the $H_{3}$ flux supported by the three-cycle $I'\times S^{2}$, there is also a five-form flux threading the non-contractible five-cycle which is obtained as a fibration of the $\hat{S}^{2}$ over the above three-cycle:  $(I'\times S^{2})\times\hat{S}^{2}$. The five-form here is the gauge-variant combination $\tilde{F}'_{5}= F_{5}+C_{2}\wedge H_{3}$, with $C_{2}$ being the gauge potential of the R-R three-form:

\hfill\break
\begin{align*}
&N_{D3}^{D5}=\int_{(I\times \hat{S}^{2})\times S^{2}}F_{5}-B_{2}\wedge F_{3} \\ 
&N_{D3}^{NS5}=\int_{(I'\times S^{2})\times \hat{S}^{2}}F_{5}+C_{2}\wedge H_{3} 
\end{align*}
\hfill\break

The above D3 charges are called \textit{Page} charges. This subtlety in defining the D3 brane charge is explained in detail in \cite{Assel:2011xz}. Briefly, it turns out that in the presence of the two kinds of fivebranes, the RR five-form, which is gauge-invariant, corresponds to a non-conserved current generating a non-conserved D3-brane charge. However, a particular definition, of a gauge-variant this time, five form as a combination of the RR fiveform and the gauge potentials for the RR and NS NS three-forms, results to a local, conserved and quantized but gauge-variant D3 brane charge, which is called Page charge. in this setup, there are two ways to define such a five-form, respecting the criterion of which out of the two gauge potentials can be globally defined on each of the non-contractible five-cycles in introduced above. Apart from being supported on fivebrane singularities, D3 brane charge can also be found at the asymptotic regions of the strip.

In the figure above, there is a new piece of information regarding the geometry of the asymptotic regions. Substituting the form of the harmonic functions introduced above in the metric, results to asymptotic regions with geometry $AdS_{5}\times S^{5}$. $AdS_{5}\times S^{5}$ is the near horizon geometry of a number of D3 branes and therefore there is also charge associated with these regions. Nevertheless,  this charge will not play any role, given that the focus is given to the case where these asymptotic $AdS_{5}\times S^{5}$ regions are capped-off.
 The limit of interest in the presented solutions is the one where $\alpha,\hat{\alpha}\rightarrow\infty$, which is smooth. In this limit the asymptotic $AdS_{5}\times S^{5}$ regions cap-off and the points at infinity become regular interior points of the full ten-dimensional geometry. This has been given in detail in \cite{Assel:2011xz} The harmonic functions that describe this setting are introduced below:
 
 \begin{equation}
h=-\sum_{i=1}^{q}\gamma_{i}\text{log}\Big[\text{tanh}\big(\frac{i\pi}{4}-\frac{z-\delta_{i}}{2}\big)\Big]+\text{c.c},\quad
\hat{h}=-\sum_{j=1}^{\hat{q}}\hat{\gamma}_{j}\text{log}\Big[\text{tanh}\big(\frac{z-\hat{\delta}_{j}}{2}\big)\Big]+\text{c.c}
\end{equation}

\begin{center}
\begin{figure}[t!]
\centering
\includegraphics[width=.66\textwidth,scale=0.6,clip=true]{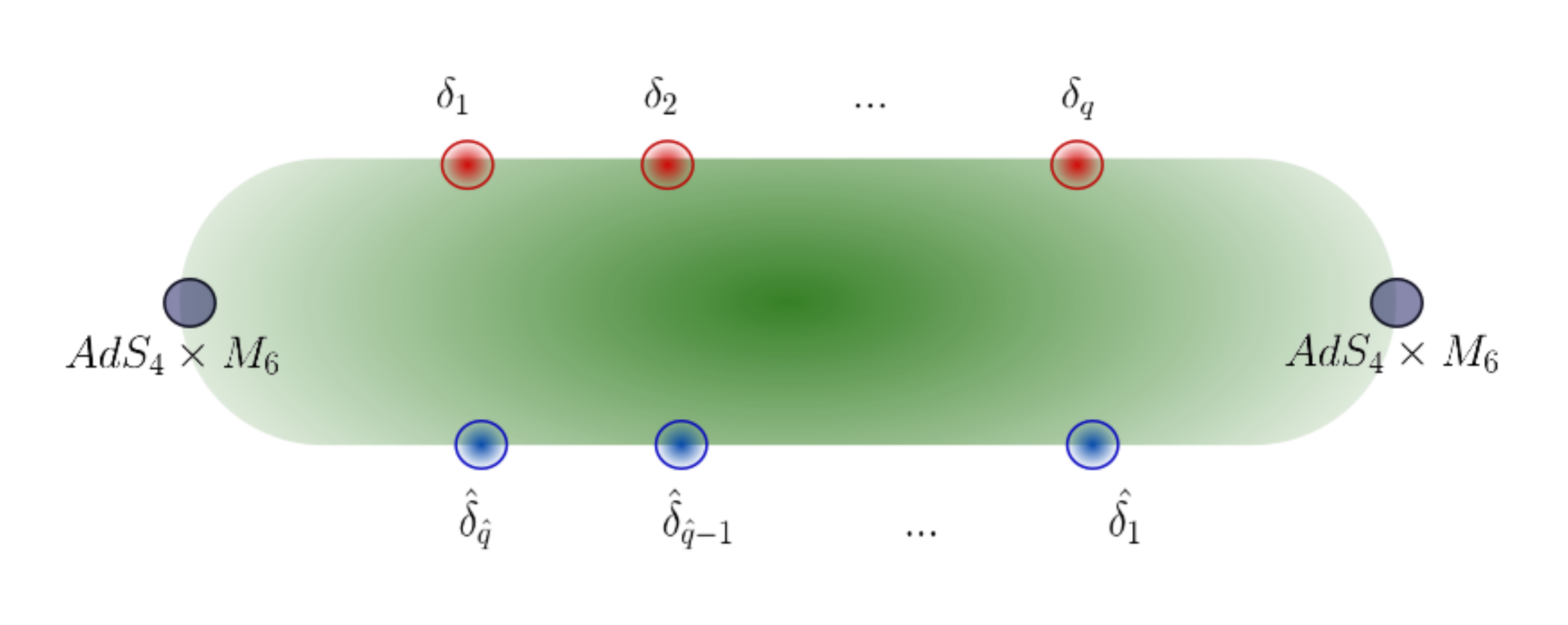}
\caption{Compactification limit: As $\alpha,\hat{\alpha}\rightarrow\infty$, the asymptotic regions close and are replaced by regions homeomorphic to $AdS_{4}\times M_{6}$. Here, solutions with multiple fivebrane singularities are depicted.}
\end{figure}
\end{center}

, where for the positions of the singularities have opposite ordering. This choice of harmonic functions corresponds to the compact geometry $AdS_{4}\times M_{6}$ as it can be verified by substituting them into the metric ansatz. Regarding the fivebrane charges, we still have that the number of D5 (NS5) branes at each stack is proportional (or equal, depending on normalization) to the corresponding $\gamma_{i}$ ($\hat{\gamma}_{j}$)-parameter of the solution: $N_{D5}^{(i)}\sim\gamma_{i}$ and $N_{NS5}^{(j)}\sim\hat{\gamma}_{j}$. Finally, the D3 brane charge is obtained by the expressions for the page charges defined above, for which the substitution of the two harmonic functions gives:

\hfill\break
\begin{equation}
N_{D3}^{(i)}=N_{D5}^{(i)}\sum_{j=1}^{\hat{q}}N_{NS5}^{(j)}\frac{2}{\pi}\textbf{arctan}(e^{\hat{\delta}_{j}-\delta_{i}}),\quad N_{D3}^{(j)}=N_{NS5}^{(j)}\sum_{i=1}^{q}N_{D5}^{(i)}\frac{2}{\pi}\textbf{arctan}(e^{\hat{\delta}_{j}-\delta_{i}})
\end{equation}
\hfill\break

These expressions show that the particular solutions have the description of near horizon geometries of a brane configuration involving D3 branes being suspended between stacks of NS5 branes and D5 branes.

\begin{flushleft}
\textbf{B: The supersymmetric Janus solution}
\end{flushleft}

\hfill\break
We have introduced the pair of harmonic functions $h, \hat{h}$,  which gave a solution of the supergravity equations on $AdS_{4}\times M_{6}$ with $AdS_{5}\times S^{5}$ aymptotic regions and point singularities supporting fivebrane and threebrane charge. Here we will present the pure $AdS_{5}\times S^{5}$ solution and its Janus generalization.

Setting $\gamma=\hat{\gamma}=0$ along with $\beta=\hat{\beta}=0$, results to the simple form for the harmonic functions:

\hfill\break
\begin{align}
&h=-i\alpha\text{sinh}(z)+\text{c.c} \\
&\hat{h}=\hat{\alpha}\text{cosh}(z)+\text{c.c}
\end{align}
\hfill\break

This solution is characterized by the absence of fivebrane charges as well as by a constant dilaton ($\delta\phi=0$) and a non-vanishing five-form background. Substituting $h,\hat{h}$ in the metric ansatz by computing the various metric factors of the solution, results to the $AdS_{5}\times S^{5}$ metric:

\hfill\break
\begin{equation}
ds^{2}_{(10)}=L_{5}^{2}\Big(\text{cosh}^{2}(x)ds^{2}_{AdS_{4}}+(dx^{2}+dy^{2})+\text{sin}^{2}(y)ds^{2}_{S^{2}}+\text{cos}^{2}(y)ds^{2}_{\hat{S}^{2}}\Big)
\end{equation}
\hfill\break

Its radius is given in terms of the only parameters of the solution, $L_{5}=2(\alpha\hat{\alpha})^{1/4}$.

Reintroducing the parameters $\beta,\hat{\beta}$ in the above solution, accounts for the deformation which results to its so-called  \textit{supersymmetric Janus} generalization:

\hfill\break
\begin{align}
&h=-i\alpha\text{sinh}(z-\beta)+\text{c.c} \\
&\hat{h}=\hat{\alpha}\text{cosh}(z-\hat{\beta})+\text{c.c}
\end{align}
\hfill\break

,where the deformation corresponds to a translation along the strip $\Sigma_{(2)}$ which preserves the boundary conditions of the two harmonic functions.  This is a supersymmetric domain wall between two asymptotic regions with $AdS_{5}\times S^{5}$ geometry. The dilaton interpolates between two constant values at the asymptotic regions ($\varphi_{+},\varphi_{-}$ at $\pm\infty$) and the above parameters denote the difference between these values :$\beta=-\hat{\beta=\delta\varphi/2}$.  The radius of the Janus solution depends now on the dilaton variation, just by a rescaling by a factor $\text{cosh}^{1/4}(\delta\varphi)$: $L_{5}=2(\alpha\hat{\alpha}\text{cosh}(\delta\varphi))^{1/4}$, whereas the dilaton in the two asymptotic regions is given by $e^{2\varphi}=\hat{\alpha}e^{\pm\delta\varphi}/\alpha $.

 The plot at the left hand side presents the dilaton, plotted for different asymptotic values, as a function of the horizontal strip coordinate. The right hand side plot, is the one of the $AdS_{4}$ warp factor for the set of dilaton variations of the left plot ($\delta\varphi=0,1,2$ for the blue, magenta and purple curves accordingly). Both plots are along $y=\pi/4$. As the dilaton variation increases, the warp factor broadens and flattens. Finally, in the limit of large dilaton variation, $\delta\varphi\rightarrow$ the geometry approaches $AdS_{4}\times\mathbb{R}$.

\hfill\break
\begin{center}
\includegraphics[width=1.0\textwidth,scale=0.6,clip=true]{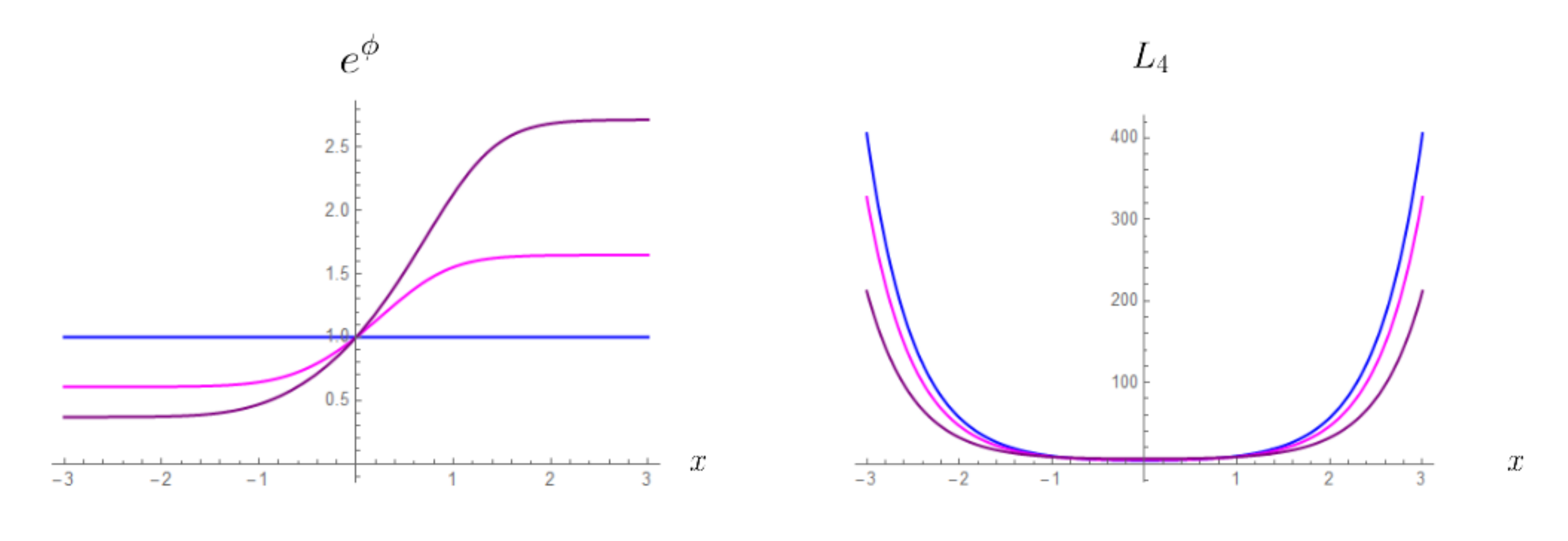}
\end{center}  
\hfill\break